\documentclass[10pt]{amsart}
\usepackage{amsmath,amsthm,amssymb,amscd}
\usepackage{epsfig}
\usepackage[margin=1.5in]{geometry}

\begin{document}

\newcommand{\Cir} {{S_3}}                                    
\newcommand{\Aff} {{\rm Aff}(\R)}                            

\newcommand{\Bc} [2] {\ensuremath{B_{#2}\langle{#1}\rangle}}  
\newcommand{\B}  [2] {\ensuremath{B_{#2}({#1})}}              
\newcommand{\BC} [2] {\ensuremath{B_{#2}[{#1}]}}              
\newcommand{\F}  [2] {\ensuremath{C_{#2}({#1})}}              
\newcommand{\FC} [2] {\ensuremath{C_{#2}[{#1}]}}              
\newcommand{\FA} [2] {\ensuremath{C_{#2}\langle{#1}\rangle}}  

\newcommand{\M}  [1] {\ensuremath{{\overline{\mathcal M}}{^{#1}_0(\R)}}}   
\newcommand{\cM} [1] {\ensuremath{{\mathcal M}_{0}^{#1}}}                  
\newcommand{\CM} [1] {\ensuremath{{\overline{\mathcal M}}{^{#1}_0}}}       
\newcommand{\oM} [1] {\ensuremath{{\mathcal M}_{0}^{#1}(\R)}}              
\newcommand{\oZ} [1] {\ensuremath{{\mathcal Z}^{#1}}}                      

\newcommand{\PV}  [1] {\ensuremath{\Pj V^{#1}}}       
\newcommand{\SV}  [1] {\ensuremath{\Sg V^{#1}}}
\newcommand{\PVH} [1] {\ensuremath{\Pj V^{#1}_{\mathcal H}}}
\newcommand{\PVM} [1] {\ensuremath{\Pj V^{#1}_{\#}}}
\newcommand{\Diag} {\Delta_{\mathcal M}}

\newcommand{\tr} {\mathcal T}
\newcommand{\cod} {{\mathfrak {X}}}
\newcommand{\m}{{\mathfrak m}}                             

\newcommand{\PGLC} {\Pj\Gl_2(\C)}                          
\newcommand{\CP} {\C\Pj^1}                                 
\newcommand{\PGL} {\Pj\Gl_2(\R)}                           
\newcommand{\RP} {\R\Pj^1}                                 
\newcommand{\C} {{\mathbb C}}                              
\newcommand{\R} {{\mathbb R}}                              
\newcommand{\I} {{\mathbb I}}                              
\newcommand{\Z} {{\mathbb Z}}                              
\newcommand{\Pj} {{\mathbb P}}                             
\newcommand{\T} {{\mathbb T}}                              
\newcommand{\Sg} {{\mathbb S}}                             
\newcommand{\Gl} {{\rm Gl}}                                

\theoremstyle{plain}
\newtheorem{thm}{Theorem}
\newtheorem{prop}[thm]{Proposition}
\newtheorem{cor}[thm]{Corollary}
\newtheorem{lem}[thm]{Lemma}
\newtheorem{conj}[thm]{Conjecture}
\newtheorem{construction}{Construction}

\theoremstyle{definition}
\newtheorem{defn}[thm]{Definition}
\newtheorem{exmp}[thm]{Example}

\theoremstyle{remark}
\newtheorem*{rem}{Remark}
\newtheorem*{hnote}{Historical Note}
\newtheorem*{nota}{Notation}
\newtheorem*{ack}{Acknowledgments}

\title {Combinatorial equivalence of real moduli spaces}

\author{Satyan L.\ Devadoss}
\address{Department of Mathematics and Statistics, Williams College, Williamstown, MA 01267}
\email{satyan.devadoss@williams.edu}
\thanks{The author was partially supported by NSF grant DMS-0310354.}

\maketitle


\baselineskip=15pt

%
%
\section*{Introduction}

The Riemann moduli space ${\mathcal M}_g^n$ of surfaces of genus
$g$ with $n$ marked points has become a central object in
mathematical physics.  Its importance was emphasized by
Grothendieck in his famous \emph{Esquisse d'un programme}. The
special case \cM{n} is a building block leading to higher genera,
playing a crucial role in the theory of Gromov-Witten invariants,
symplectic geometry, and quantum cohomology.  There is a
Deligne-Knudsen-Mumford compactification \CM{n} of this space
coming from Geometric Invariant Theory which allows
collisions of points of the configuration space.  This
description comes from the repulsive potential observed by
quantum physics:  Pushing particles together creates a spherical
bubble onto which the particles escape \cite{pw}.  In other
words, as points try to collide, the result is a new bubble fused
to the old at the point of collision where the collided points
are now on the new bubble.  The phenomena is dubbed as {\em
bubbling}; the resulting structure is called a {\em bubble-tree}.

Our work is motivated by the {\em real} points \M{n} of this
space, the set of points fixed under complex conjugation. These real moduli spaces have importance in their own right,
beginning to appear in many areas.
For instance, Goncharov and Manin \cite{gm} recently introduce \M{n} in discussing
$\zeta$-motives and the geometry of \CM{n}.

The real spaces, unlike their complex counterparts, have a tiling
that is inherently present in them.  This allows one to
understand and visualize them using tools ranging from
arrangements, to reflection groups, to combinatorics. This
article began in order to understand why the two pictures in
Figure~\ref{pv3m06} are the same:  Both of them have identical
cellulation, tiled by 60 polyhedra known as associahedra.  It was
Kapranov who first noticed this relationship, relating \M{n} to
the braid arrangement of hyperplanes.  We provide an intuitive,
combinatorial formulation of \M{n} in order to show the
equivalence in the figure.  Along the way, we provide a
construction of the associahedron from truncations of certain
products of simplices.

A configuration space of $n$ ordered, distinct particles on a
manifold $M$ is defined as
$$\F{M}{n} = M^n - \Delta, \ \ \ {\rm
where} \ \Delta = \{(x_1, \ldots , x_n) \in M^n \ | \ \exists \
i,j, \ x_i = x_j \}.$$ 
The recent work in physics around conformal field theories has
led to an increased interest in the configuration space of $n$
labeled points on the projective line.  The focus is on a
quotient of this space by $\PGLC$, the affine automorphisms on
$\CP$.  The resulting variety \cM{n} is the moduli space of
Riemann spheres with $n$ labeled punctures.

\begin{defn}
The {\em real} moduli space of $n$-punctured Riemann spheres is
$$\oM{n} \ = \ \F{\RP}{n}/\PGL,$$
where $\PGL$ sends three of the points to $0, 1, \infty$.
\end{defn}

\noindent This moduli space encapsulates the new constructions of
the associahedra developed below.

%
%
\section*{The Simplex}

\subsection*{}
For a given manifold $M$, the symmetric group $\Sg_n$ acts freely on the configuration space
\F{M}{n} by permuting the coordinates, and the quotient manifold
$\B{M}{n} \, = \, \F{M}{n} / \, \Sg_n$ is the space of $n$
unordered, distinct particles on $M$.  The closure of this space
in the product is denoted by \Bc{M}{n}.  Let $\Aff$ be the group
of affine transformations of $\R$ generated by translating and
scaling.  The space $\B{\R}{n+2}/\Aff$ is the open $n$-simplex:
The leftmost of the $n+2$ particles in $\R$ is translated to $0$
and the rightmost is dilated to $1$, and we have the subset of
$\R^n$ where
\begin{equation}
0 \ < \ x_ 1 \ < \ x_2 \ < \ \cdots \ < \ x_{n-1} \ < \ x_{n} \ < \ 1.
\label{e:simplex}
\end{equation}
The closure of this space is the $n$-simplex $\Delta_n$ whose
codimension $k$ face can be identified by the set of points with
exactly $k$ equalities of (\ref{e:simplex}).

\begin{figure} [h]
\centering {\includegraphics {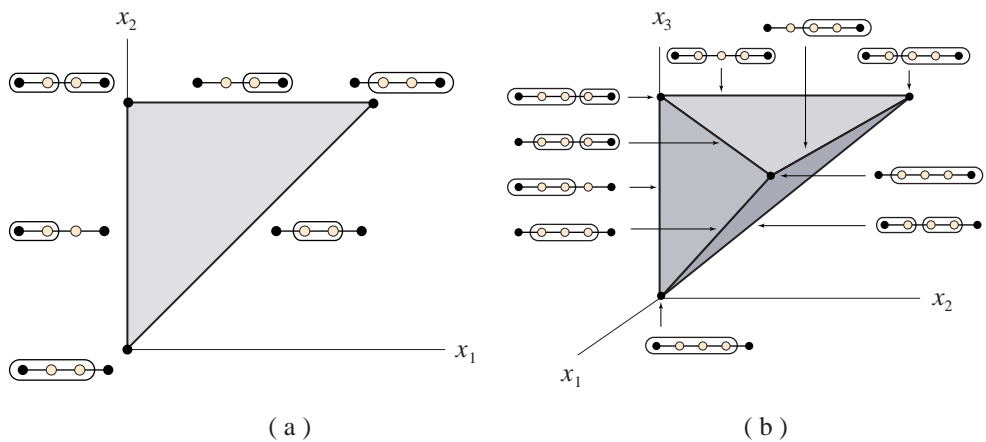}}
\caption{Labeling of vertices and edges of $\Delta_2$ and $\Delta_3$.}
\label{simplex23p}
\end{figure}

\begin{nota}
If we let $I_2$ denote the unit interval $[0, 1] \subset \R$ with
{\em fixed} particles at the two endpoints, then the $n$-simplex
can be viewed as the closure \Bc{I_2}{n}.  We use bracket
notation to display this visually:  Denote the $n$ particles on
the interval $I_2$ as nodes on a path, with the fixed ones as
nodes shaded black.  When the inequalities of (\ref{e:simplex})
become equalities, draw brackets around the nodes representing
the set of equal points on the interval.  For example,
\includegraphics{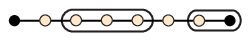} corresponds to the configuration
$$0 \ < \ x_1 \ < \ x_2 \ = \  x_3 \ = \ x_4 \ < \ x_5 \ < \ x_6 \ = \ 1.$$
We call such a diagram a {\em bracketing}.
Figure~\ref{simplex23p} depicts $\Delta_2$ and $\Delta_3$ along
with a labeling of vertices and edges.
\end{nota}

\subsection*{}
The associahedron is a convex polytope originally defined
by Stasheff~\cite{sta} for use in homotopy theory in connection with
associativity properties of $H$-spaces.  It continues to appear
in a vast number of mathematical fields,
currently leading to numerous generalizations.

\begin{defn}
Let ${\mathfrak A}(n)$ be the poset of bracketings of a path with
$n$ nodes, ordered such that $a \prec a'$ if $a$ is obtained from
$a'$ by adding new brackets.  The {\em associahedron} $K_n$ is a
convex polytope of dimension $n-2$ whose face poset is isomorphic
to ${\mathfrak A}(n)$.
\end{defn}

\begin{figure}[h]
\includegraphics {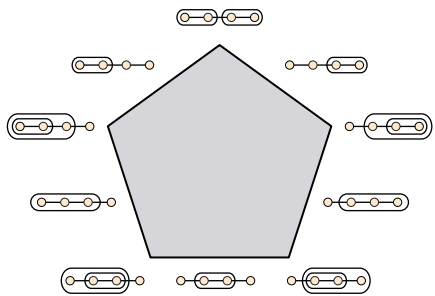}
\caption{Associahedron $K_4$.}
\label{k4p}
\end{figure}

\begin{exmp}
Figure~\ref{k4p} shows the two-dimensional $K_4$ as the pentagon.
Each edge of $K_4$ has one set of brackets, whereas each vertex
has two.  Figure~\ref{k4k5p}(b) depicts $K_5$ with only the {\em
facets} (codimension one faces) labeled here.
\end{exmp}

\begin{figure}[h]
\includegraphics {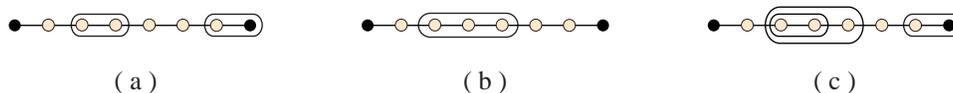}
\caption{Compatibility of bracketings.}
\label{union}
\end{figure}

\noindent Two bracketings are {\em compatible} if the brackets of
the superimposition do not intersect.  Figure~\ref{union} shows an
example of two compatible bracketings (a) and (b).  It follows
from the definition of $K_n$ that two faces are adjacent if and
only if their bracketings are compatible.  Furthermore, the face
of intersection is labeled by the superimposed image (c).

\subsection*{}
A well-known construction of the associahedron from the simplex
via truncating hyperplanes is given in the Appendix of
\cite{sta2}.
A reformulation from the perspective of configuration spaces is as follows:

\begin{rem}
An $n$-polytope is {\em simple} if every $k$-face is contained in $n-k$ facets.  The simplex is a simple polytope and a truncation of a simple polytope remains simple.
\end{rem}

\begin{construction}
Choose the collection $\mathcal C$ of codimension $k$ faces of
the $n$-simplex \Bc{I_2}{n} which correspond to configurations
where $k+1$ {\em adjacent} particles collide. Truncating elements
of \ $\mathcal C$ in increasing order of dimension results in
$K_{n+2}$. 
\label{t:simplexkn}
\end{construction}

\begin{proof}
We show the construction to be well defined, that truncation is a commutative operation for faces of the same dimension.  In other words, if two codimension $k$ faces $F_1$ and $F_2$ of $\mathcal C$ intersect at a codimension $(k+1)$ face $G$, then $G$ is in $\mathcal C$.  Indeed, this is an immediate consequence of what it means to be an element of $\mathcal C$:  Since $F_1$ and $F_2$ each have $k$ adjacent equalities in (\ref{e:simplex}), then $G$ must have $k+1$ {\em adjacent} equalities since $G = F_1 \ \cap \ F_2$.

We show that the face poset of $\Delta_n$, as faces in $\mathcal C$ are truncated, changes to the face poset of $K_{n+2}$.  Let $F$ be a codimension $k$ face in $\mathcal C$ and $\mathcal K_F$ be the collection of faces of the polytope that intersect $F$.  By definition of truncation, there exists a bijection $\phi: Y_F \rightarrow \mathcal K_F$ between the faces of $Y_K$ to elements in $\mathcal K_F$.  Label each face $f$ of $Y_F$ with the superimposition of the bracket labelings of $F$ and $\phi(f)$.  It is clear the labelings of $F$ and $\phi(f)$ will be compatible from the adjacency relation of the faces.

Since our polytope is simple, truncating $F$ replaces it with a facet $Y_F = F \times \Delta_{k-1}$. Since $F$ is defined by $k+1$ adjacent particles colliding, the simplex $\Delta_{k-1}$ introduced in the truncation inherits the bracket labeling of \Bc{I_2}{k+1}.  Indeed, we are not allowing the $k+1$ particles to collide at once, but resolving all possible orderings in which the collisions could occur.  After iterating this procedure over all elements of $\mathcal C$, the face poset of the resulting polytope will isomorphic to $K_{n+2}$.
\end{proof}

\begin{rem}
A proof of this construction using face posets and bracketings in
a general context of graphs is given in \cite[\S5]{small}.
\end{rem}

\begin{figure} [h]
\centering {\includegraphics {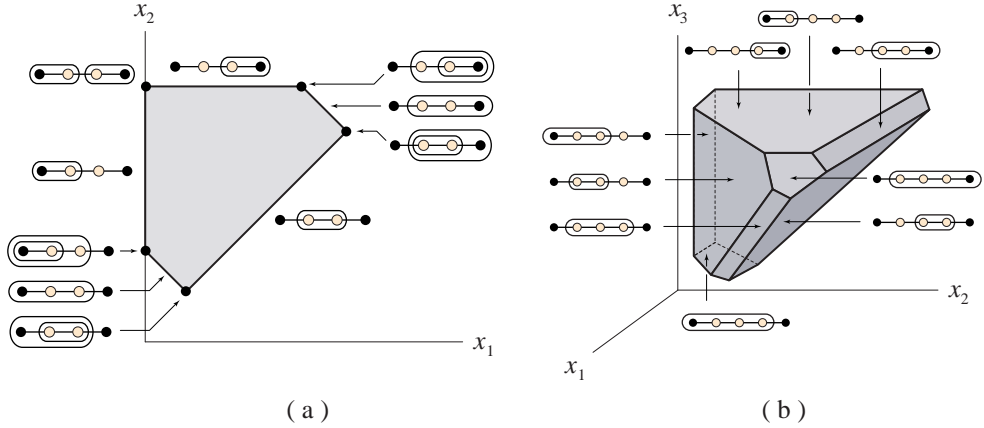}}
\caption{(a) Vertices and edges of $K_4$ labeled. \ (b) Facets of $K_5$ labeled.}
\label{k4k5p}
\end{figure}

\begin{exmp}
Figure~\ref{k4k5p}(a) shows $K_4$ after truncating the two
vertices \includegraphics{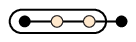} and  \break
\includegraphics{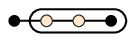}
of $\Delta_2$ given in Figure~\ref{simplex23p}.  Each vertex is
now replaced by a facet given the same labeling as the original
vertices.  However, the new vertices introduced by shaving are
labeled with nested parentheses, seen as the superimposition of
the respective diagrams.  Similarly, Figure~\ref{k4k5p}(b) displays
$K_5$ with facets diagrams after first shaving two vertices and
then three edges of $\Delta_3$.  Compare this with
Figure~\ref{simplex23p}.
\end{exmp}

This construction of $K_n$ from the simplex is the real 
{\em Fulton-MacPherson} \cite{fm} compactification of the configuration space \B{I_2}{n}.  We
denote this as \BC{I_2}{n}.   Casually speaking, one is not only interested in when $k$
adjacent particles collide, but in resolving that singularity by
ordering the collisions.  For example,
\raisebox{-.9mm}{\includegraphics{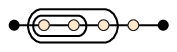}} not just conveys
that the three particles have collided, but that the first two
particles collided before meeting with the third.

\begin{rem}
In the original closed simplex, the number of equalities
(collisions) correspond to the codimension of the cell.  After
the compactification, the codimension is given by the number of
brackets.
\end{rem}

%
%
\section*{Products of Simplices}

\subsection*{}
We extend the notions above to triple products of simplices.  In
doing so, we see new combinatorial constructions of the
associahedron.  Let $\Cir$ denote a circle with three distinct
fixed particles. The space \Bc{S_3}{n} is combinatorially
equivalent to the product of three simplices $\Delta_i \times
\Delta_{j} \times \Delta_{k}$, with $i + j + k = n$.  Indeed, the
different types of simplicial products depend on how the $n$
particles are partitioned among the three regions, each region
defined between two fixed particles.  Note that each
configuration of $k$ particles which fall between two fixed
particles give rise to the $k$-simplex \Bc{I_2}{k}.

\begin{figure}[h]
\includegraphics {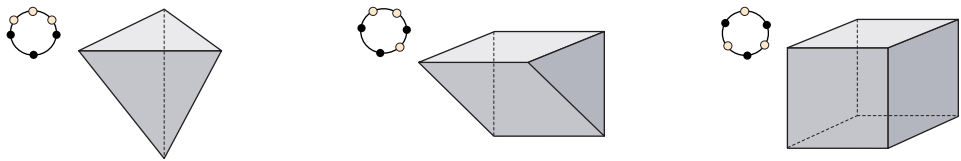}
\caption{Three types of simplicial products with three particles.}
\label{F3-simplex}
\end{figure}

\begin{exmp}
There are three possibilities when $n=3$:  The simplex
$\Delta_3$, the prism $\Delta_2 \times \Delta_1$, and the cube
$\Delta_1 \times \Delta_1 \times \Delta_1$ as presented in
Figure~\ref{F3-simplex}.
\end{exmp}

\begin{construction}
Let ${\mathfrak B}(n)$ be the poset of bracketings of $\Cir$ with
$n - 2$ additional nodes \break partitioned into the three
regions, where no bracket contains more than one of the three
marked nodes of $\Cir$.  Order them such that $b \prec b'$ if $b$
is obtained from $b'$ by adding new brackets.  The face poset \ ${\mathfrak A}(n)$ of \
$K_n$ is isomorphic to ${\mathfrak B}(n)$.
\label{l:bn}
\end{construction}

\begin{figure}[h]
\includegraphics {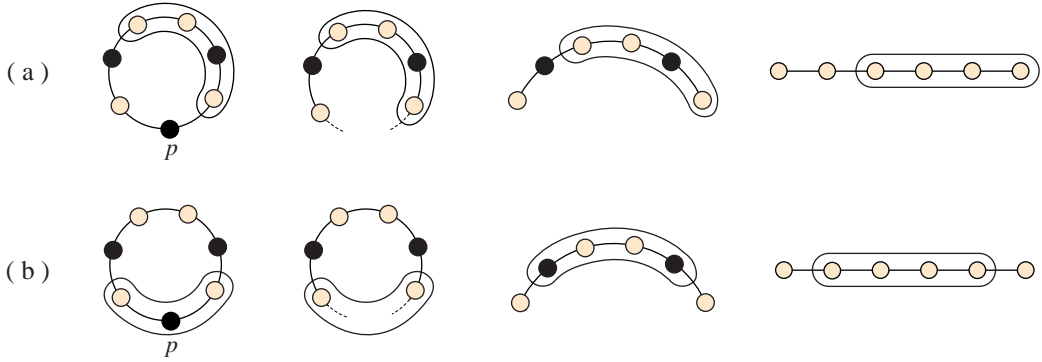}
\caption{Bijection from ${\mathfrak B}(n)$ to ${\mathfrak A}(n)$.}
\label{proof}
\end{figure}

Choose any one of the three fixed particles
of $\Cir$, call it $p$.  The particles of $\Cir - p$ can be
viewed as $n$ particles on the line.  If a bracket does not
contain $p$, preserve this bracketing on the line;  see
Figure~\ref{proof}(a).  If a bracket does contain $p$, choose the
bracket on the line that encloses the complementary set of
particles; see Figure~\ref{proof}(b). This is a bijection of
posets since a bracket on $\Cir$ can contain at most one fixed
particle.

\begin{rem}
Each partition of the $n - 2$ nodes in $\Cir$ gives rise to a
different poset that is isomorphic to ${\mathfrak A}(n)$.
\end{rem}


\subsection*{}
We look at the compactification \BC{\Cir}{n}.  Analogous to
Construction~\ref{t:simplexkn}, we specify certain faces of
$\Delta_x \times \Delta_y \times \Delta_z$ to be truncated,
namely the codimension $k$ faces where $k+1$ adjacent particles
collide. Indeed, each facet of the polytope \BC{\Cir}{n} will
correspond to a unique way of adding a bracket around the $n + 3$
particles ($n$ free and 3 fixed) in $\Cir$.  The restriction will
be that no bracket will include more than one fixed particle, for
this would imply that the fixed particles inside the bracket
would be identified.

\begin{figure}[h]
\includegraphics {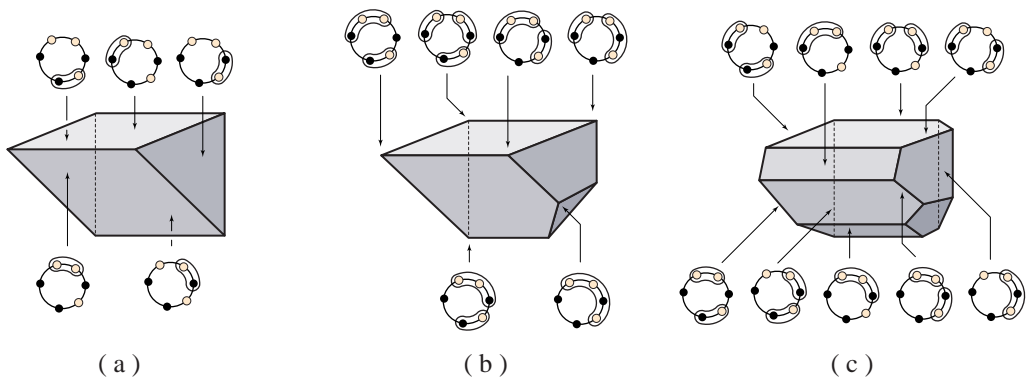}
\caption{Truncation and labeling of $\Delta_2 \times \Delta_1$.}
\label{Prism-k5}
\end{figure}

\begin{exmp}
Figure~\ref{Prism-k5}(a) shows the prism in
Figure~\ref{F3-simplex} with labeling of the top dimensional
faces.    Figure~\ref{Prism-k5}(b) shows the labeling of the
vertices, along with the new facet obtained by shaving a vertex
(codimension three) where four adjacent particles collide.
Similarly, part (c) is the labeling of the edges, along with the
truncation of three of them.  Notice that the resulting polytope
is combinatorially equivalent to $K_5$.
\end{exmp}

\begin{construction}
Choose the collection of codimension $k$ faces of $\Delta_x
\times \Delta_y \times \Delta_z$ which correspond to
configurations where $k+1$ {\em adjacent} particles collide.
Truncating elements of this collection in increasing order of
dimension results in $K_{x+y+z+2}$. 
\label{t:3simplexkn}
\end{construction}

Since $\Delta_x \times \Delta_y \times \Delta_z$ is simple,
truncating a codimension $k$ face $F$ replaces it with
a  product $F \times \Delta_{k-1}$.  Label the faces of $F \times
\Delta_{k-1}$ with superimposition of neighboring faces.
Truncating all elements produces a face poset structure
isomorphic to ${\mathfrak B}(n)$.  Then use
Construction~\ref{l:bn}.

\begin{cor}
Let $p_k(n)$ be partitions of $n$ into exactly $k$ parts.  There are
$$p_3(n-3) \ + \ p_2(n-2) \ + \ 1$$
different ways of obtaining $K_n$ from iterated truncations of simplicial products.
\end{cor}

\begin{figure}[h]
\includegraphics {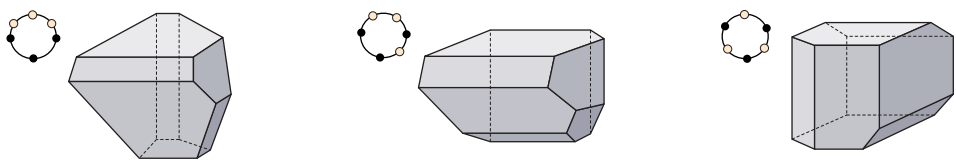}
\caption{Iterated truncations of polytopes resulting in $K_5$.}
\label{F3-k5}
\end{figure}

\begin{figure}[h]
\includegraphics {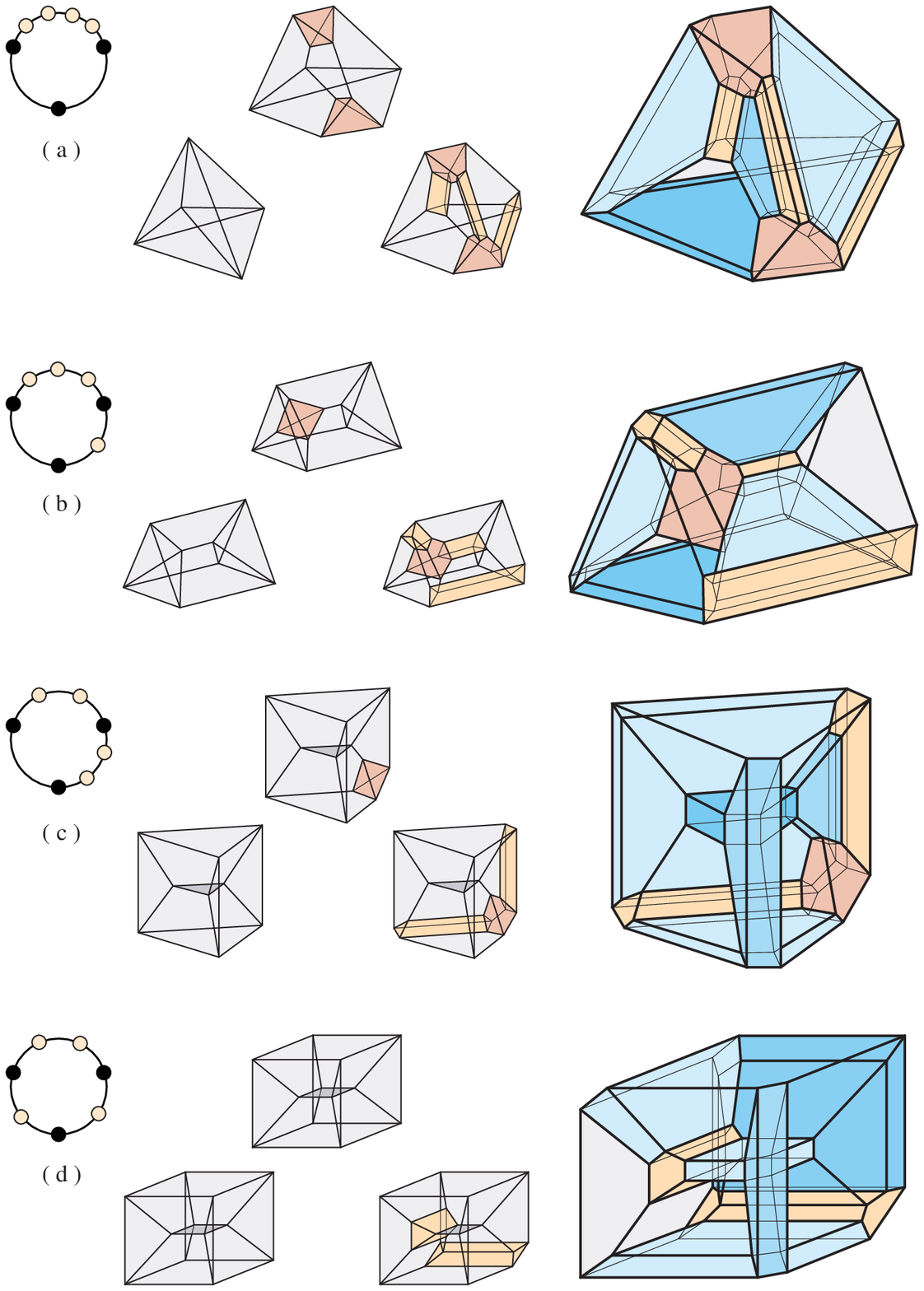}
\caption{Schlegel diagrams of the iterated truncations of $4$-polytopes resulting in $K_6$.}
\label{F4-k6}
\end{figure}

Indeed, for each triple product of simplices, there exists a
method to obtain the associahedron from iterated truncations of
faces.  Figure~\ref{F3-k5} shows $K_5$ from truncations
of the three polytopes in  Figure~\ref{F3-simplex}.
Figure~\ref{F4-k6} displays the {\em Schlegel diagrams} of
four $4$-polytopes, the (a) $4$-simplex, (b) tetrahedral prism,
(c) product of triangles, and (d) product of triangle and
square.  Each is truncated to (combinatorial equivalent) $K_6$
associahedra, each with seven $K_5$ and seven pentagonal prism
facets.

%
%
\section*{The Braid Arrangement}

\subsection*{}
We relate the combinatorial structure of the associahedron to a
tiling of spaces.  This yields an elegant framework for associating
Coxeter complexes to certain moduli spaces.  We begin
with some background \cite{bro}.  The symmetric
group $\Sg_{n+2}$ is a finite reflection group acting on
$\R^{n+2}$ as reflections $(ij)$ across the hyperplanes $\{x_i =
x_j\}$, forming the {\em braid arrangement} of hyperplanes
$\mathcal H$.  The {\em essential} subspace under the action of
$\Sg_{n+2}$ is the hyperplane $V^{n+1}$ defined by $\Sigma x_i =
0$.  This space is tiled by simplicial cones, defined by $n+1$
inequalities
\begin{equation}
x_{i_1} \ \leq \ x_{i_2} \ \leq \ \cdots \ \leq \ x_{i_{n+1}} \ \leq \ x_{i_{n+2}}.
\label{e:cones}
\end{equation}
Let \SV{n} be the sphere in $V^{n+1}$.  The braid arrangement
gives these spaces a cellular decomposition into $(n+2)!$
chambers.  Each chamber of \SV{n} is an $n$-simplex, defined by
(\ref{e:cones}) where not all inequalities are
equalities.\footnote{The point where all equalities exist is at
the cone point, which is not contained in the sphere.}

\begin{defn}
A {\em cellulation} of a manifold $M$ is formed by gluing
together polytopes using combinatorial equivalence of their
faces, together with the decomposition of $M$ into its cells.
\end{defn}

\begin{prop}
Let \FA{\R}{n} denote the closure of $\F{\R}{n}/\Aff$.  Then
\FA{\R}{n} has the same cellulation as \SV{n-2}. \label{p:consph}
\end{prop}

\begin{proof}
Let $\vec{a_1}, \ldots, \vec{a_n} \in \R^n$ such that $\vec{a_i} = -(\vec{e_1} + \cdots + \vec{e_n}) + n\vec{e_i}$.  Note that $\sum{\vec{a_i}} = 0$, $\vec{a_i}\ \in\ {\langle 1,\ldots,1\rangle}^{\bot}$, and 
\begin{equation*}
\vec{a_i} \cdot \vec{a_j} =  \left \{ \begin{array}{ll}
n^2 - n & \textrm{for $i=j$} \\
-n & \textrm{for  $i \neq j$.}
\end{array} \right.
\end{equation*}
Let $v = \langle v_1,\ldots, v_n \rangle \in \FA{\R}{n}$.  Define the map $\varphi: \FA{\R}{n+2} \rightarrow \SV{n-2}$ such that
$$\varphi(v) = \frac{\sum v_i\vec{a_i}}{\vert \sum v_i\vec{a_i} \vert}.$$
An ordering of the $n$ points $v_1 \leq \cdots \leq v_n$ defines a chamber in $\FA{\R}{n}$.  Similarly, a chamber of \SV{n-2} corresponds to an ordering of elements as in equation (\ref{e:cones}).
We show that $\varphi(v_1) \leq \cdots \leq \varphi(v_n)$.  For each $v_i \leq v_j$, 
$$\vec{a}_j \cdot \varphi{(v)} \ - \ \vec{a}_i \cdot \varphi{(v)} \ = \  n^2(v_i - v_j) \ \geq \ 0.$$
Now $\sum\varphi(v_i) = 0$ \ since \ $\varphi(v_i) \in \Sg V^{n-2}$, so 
$$\vec{a}_i \cdot \varphi{(v)} =  -(\varphi(v_1) + \cdots + \varphi(v_n)) + n\varphi(v_i) =  n\varphi{(v)}_i,$$
and thus $\varphi(v_i) \leq \varphi(v_j)$ preserving the chamber structure.   It is easy to show that $\varphi$ is a homeomorphism. 
Since a codimension $k$ face of both spaces is where exactly $k$
equalities in $\langle v_1,\ldots, v_n \rangle$ occur, the cellulation naturally follows.
\end{proof}

\begin{figure} [h]
\centering {\includegraphics {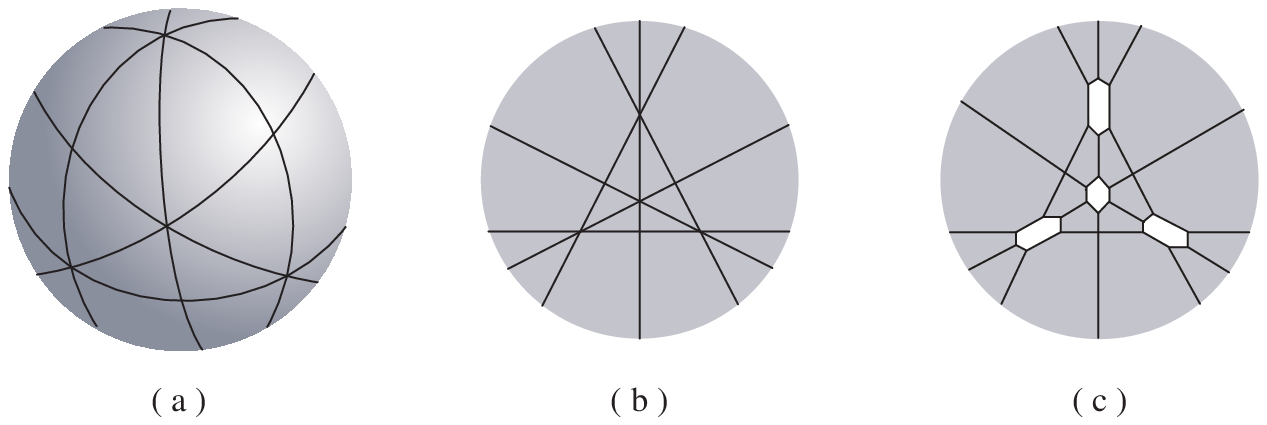}}
\caption{(a) \SV{2},\ (b) \PV{2} and (c) \PVM{2}.}
\label{svpvm05}
\end{figure}

Indeed, each simplicial chamber of \SV{n} corresponds to an
arrangement of $n+2$ particles on an interval, resulting in
\Bc{I_2}{n}.  A chamber of \PV{n}, the projective sphere in
$V^{n+1}$, identifies two antipodal chambers of \SV{n}.
Figures~\ref{svpvm05}(a) and \ref{svpvm05}(b) depict the $n=2$
case. Observe that quotienting by translations of $\Aff$ removes
the inessential component of the arrangement, scaling (by a
factor of $s \in \R^+$) pertains to intersecting $V^n$ with the
sphere, and {\em dilating} (by a factor of $s \in \R^*$) results
is \PV{n}.

\subsection*{}
The collection of hyperplanes $\{x_i = 0 \ | \ i = 1, \ldots,
n\}$ of $\R^n$ generates the {\em coordinate} arrangement. Let
$M$ be a manifold and $D \subset M$ a union of codimension one
submanifolds which dissects $M$ into convex polytopes.
  A crossing (of $D$) in $M$ is {\em normal}\, if it
is locally isomorphic to a coordinate arrangement.  If every
crossing is normal, then $M$ is {\em right angled}.  An operation
which transforms any crossing into a normal crossing
involves the algebro-geometric concept of a blow-up.

\begin{defn}
For a linear subspace $X$ of a vector space $Y$, we {\em blow up
\,$\Pj Y$ along \,$\Pj X$} by removing $\Pj X$, replacing it with
the sphere bundle associated to the normal bundle of $\Pj X
\subset \Pj Y$, and then projectifying the bundle.
\end{defn}

Blowing up a subspace of a cell complex truncates faces of
polytopes adjacent to the subspace.  As mentioned above with
truncations, a general collection of blow-ups is usually
non-commutative in nature; in other words, the order in which
spaces are blown up is important.  For a given arrangement, De
Concini and Procesi \cite{dp} establish the existence (and
uniqueness) of a {\em minimal building set}, a collection of
subspaces for which blow-ups commute for a given dimension, and
for which the resulting space is right angled.

For an arrangement of hyperplanes, the method developed by De
Concini and Procesi compactifies their complements by iterated
blow-ups of the minimal building set.  In the case of
the arrangement $X^n - \F{X}{n}$, their procedure yields the
Fulton-MacPherson compactification of $\F{X}{n}$. We can view
\PV{n} as a configuration space, where the codimension $k$
elements of the minimal building set are the subspaces
\begin{equation}
x_{i_1} = x_{i_2} = \cdots = x_{i_{k+1}}
\label{e:min}
\end{equation}
of \PV{n} where $k + 1$ adjacent particles collide. 
Let \PVM{n} denote the space \PV{n} after iterated blow-ups along
elements of the minimal building set in increasing order of dimension.

\begin{thm} \textup{\cite{kap}}
\PVM{n} is tiled by  $\frac{1}{2}(n+2)!$ copies of associahedra $K_{n+2}$.
\end{thm}

\noindent Indeed, this is natural since the blow-up of all codimension $k$ subspaces (\ref{e:min}) truncates the collection $\mathcal C$ of codimension $k$ faces of the simplex defined in Construction~\ref{t:simplexkn}.  Figure~\ref{svpvm05}(c) shows \PVM{2} tiled by 12 associahedra $K_4$.


\subsection*{}
A combinatorial construction of \PVM{n} is presented in
\cite{dev} by gluing faces of the $\frac{1}{2}(n+2)!$ copies
of associahedra.  Associate to each $K_{n+2}$ a path with $n + 2$
labeled nodes, with two such labelings equivalent up to
reflection.  Thus each face of an associahedron is identified
with a labeled bracketing.  A {\em twist} along a bracket
reflects all the elements within the bracket (both labeled nodes
and brackets).

\begin{thm} \textup{\cite{dev}}
Two bracketings of a path with $n + 2$ labeled nodes,
corresponding to faces of $K_{n+2}$, are identified in \PVM{n} if
there exists a sequence of twists along brackets from one diagram
to another. \label{t:twist}
\end{thm}

Each element of the minimal building set corresponds to subspaces
such as (\ref{e:min}), where blowing up the subspace seeks to
resolve the {\em order} in which collisions occur at such
intersections.  Crossing from a chamber through the blown-up cell
into its antipodal one in the arrangement (from projectifying the
bundle) corresponds to reflecting the elements $\{x_{i_1},\,
x_{i_2},\, \ldots,\, x_{i_{k+1}}\}$ in the ordering. Blowing up a
minimal cell identifies faces across the antipodal chambers, with
twisting along diagonals mimicking gluing antipodal faces after
blow-ups.

\begin{figure} [h]
\centering {\includegraphics {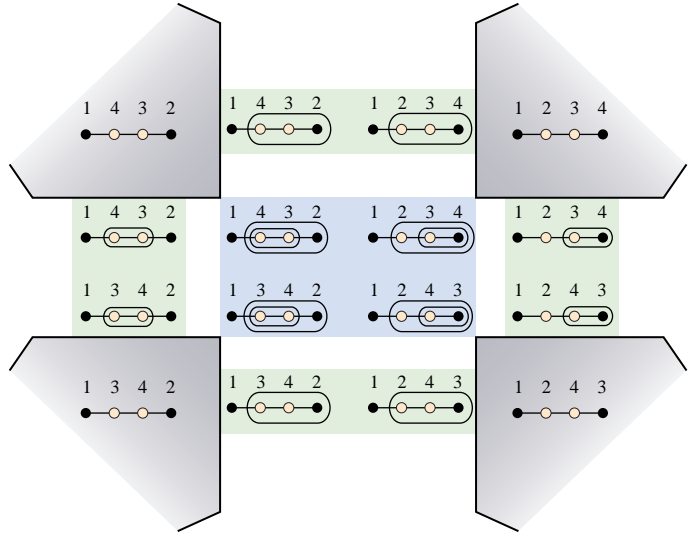}}
\caption{A local tiling of \PVM{5} displaying twisting.}
\label{m05parts}
\end{figure}

Figure~\ref{m05parts} shows a local tiling of \PVM{2} by $K_4$,
with edges (in pairs) and vertices (in fours) being identified
after twists.  Notice that after twisting a bracket containing a fixed
node, the new right- (or left-)most node becomes fixed by the
action of $\Aff$.

\begin{rem}
This immediately shows \PVM{n} to be right angled:  A codimension
$k$ face of an associahedron of \PVM{n} has $k$ brackets, with
each twist along a bracket moving to an adjacent chamber.  There
are $2^k$ such possible combinations of twists, giving a normal
crossing at each face.
\end{rem}

%
%
\section*{Kapranov's Theorem}

\subsection*{}
We start with properties of the manifold before compactification.

\begin{prop}
Let \PVH{n} denote \PV{n} minus the braid arrangement $\mathcal H$.  Then \oM{n+3} is isomorphic to \PVH{n}.
\label{p:number}
\end{prop}

\begin{proof}
Let $(x_1, \ldots, x_{n+3}) \in \F{\RP}{n+3}$.  Since a projective automorphism of $\Pj^1$ is uniquely determined by the images of three points, we can take $x_{n+1}, x_{n+2}, x_{n+3}$ to $0, 1, \infty$, respectively. Therefore,
\begin{eqnarray*}
\oM{n+3}& = & \{(x_1, \ldots, x_n) \in (\RP)^n \ | \ x_i \neq x_j, \ x_i \neq 0, 1, \infty\} \\
& = & \{(x_1, \ldots, x_n) \in (\R^1)^n \ | \ x_i \neq x_j, \ x_i \neq 0, 1\}\\
& = & \{(x_1, \ldots, x_n) \in \R^n \ | \ x_i \neq x_j, \ x_i \neq 0, 1\}.
\end{eqnarray*}
We construct a space isomorphic to \PVH{n}:  Intersect $\F{\R}{n+2}$ with the hyperplane $\{x_{n+2}=0\}$ instead of the more symmetric hyperplane $\{\Sigma x_i = 0\}$ to obtain
$$\{(x_1, \ldots, x_{n+1}) \in \R^{n+1} \ | \ x_i \neq x_j, \ x_i \neq 0\}.$$
We projectify by choosing the last coordinate to be one, resulting in
$$\{(x_1, \ldots, x_n) \in \R^n \ | \ x_i \neq x_j, \ x_i \neq 0, 1\}.$$
This is isomorphic to \PVH{n}, and the equivalence is shown.
\end{proof}

Since \oM{n+3} is isomorphic to the $n$-torus $(\RP)^n$ minus the hyperplanes $\{ x_i = x_j, \ x_i = 0, 1, \infty\}$, it follows that
$$\oM{n+3} \ =  \ \F{\Cir}{n}$$
with the three fixed points identified to $0, 1, \infty$.  As \PV{n} is tiled by simplices, the closure of \oM{n+3} is tiled by triple product of simplices, namely \Bc{\Cir}{n}. The compactification \M{n+3} is obtained by iterated blow-ups of \oM{n+3} along non-normal crossings in increasing order of dimension \cite[\S3]{sta2}.  The codimension $k$ subspaces
$$x_{i_1} = x_{i_2} = \cdots = x_{i_{k+1}}$$
and
$$x_{i_1} = x_{i_2} = \cdots = x_{i_k} = f,$$
where $f \in \{0, 1, \infty\}$, form the minimal building set, configurations where $k + 1$ adjacent particles collide on $\Cir$.  Similar to \PVM{n}, the blow-up of all minimal subspaces truncate the chambers into associahedra as defined by Construction~\ref{t:3simplexkn}.


\subsection*{}
Although the closures of \oM{n+3} and \PVH{n} are clearly
different (the torus $T^n$ and $\R\Pj^n$ respectively), Kapranov
\cite[\S4]{kap} remarkably noticed that their compactifications
are homeomorphic.\footnote{Kapranov actually proves a stronger
result for the complex analog of the statement using Chow
quotients of Grassmanians \cite{kap2}.}  We give an alternate
proof of his theorem.

\begin{thm}
\M{n+3} is homeomorphic to \PVM{n}.  Moreover, they have identical cellulation.
\end{thm}

\begin{proof}
Both \M{n+3} and \PVM{n} have the same number of chambers by
Proposition~\ref{p:number}.  Each tile of the closure of \oM{n+3}
corresponds to a triple product of simplices. Since the building
set of \oM{n+3} corresponds to the faces of \Bc{\Cir}{n} to be
truncated in Construction~\ref{t:3simplexkn}, \M{n+3} is tiled by
associahedra $K_{n+2}$, more precisely by \BC{\Cir}{n}.  We still
need to show this tiling is identical to that of \PVM{n}.

As in Theorem~\ref{t:twist}, crossing a chamber through the
blown-up cell into its antipodal one in the arrangement
corresponds to reflecting the elements within a bracket of
\BC{\Cir}{n}.  This is encapsulated by the twisting operation on
$\Cir$, similar to \PVM{n}.  Finally, Construction~\ref{l:bn} gives us
the isomorphism of cellulations between \M{n+3} and \PVM{n}.
\end{proof}

\begin{figure}[h]
\includegraphics {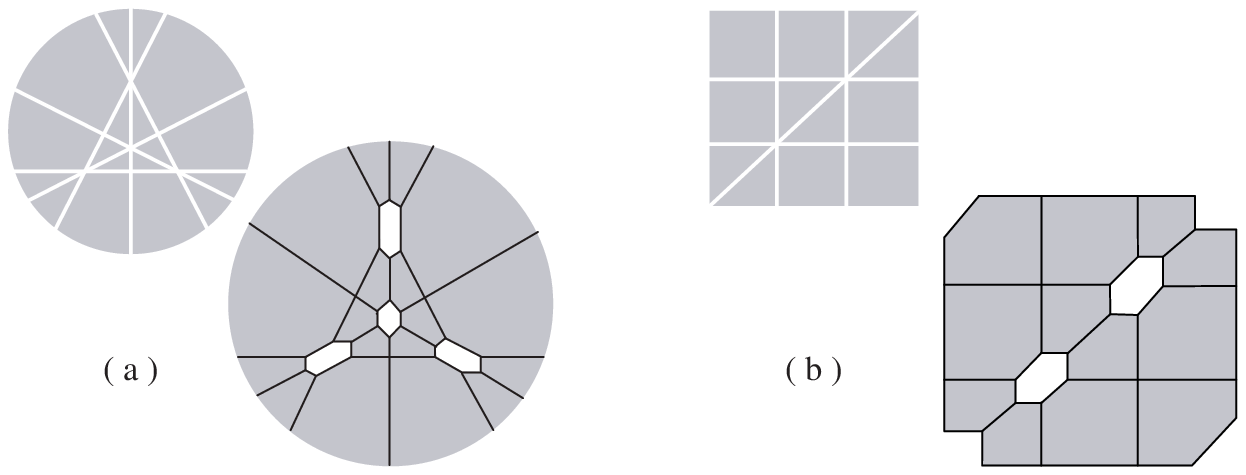}
\caption{(a) \PVH{2} and (b) \oM{5} before and after compactification.}
\label{hyperplanes2}
\end{figure}

\begin{exmp}
The top diagrams of Figure~\ref{hyperplanes2} present (a) \PVH{2}
tiled by open simplices, and (b) \oM{5}, the 2-torus minus the
hyperplanes $\{x_1 = x_2, \ x_i = 0, 1, \infty \}$ tiled by open
simplices and squares.  After minimal blow-ups, the resulting
(homeomorphic) manifolds are (a) $\#^5 \, \R\Pj^2$ and (b) $T^2
\, \#^3 \, \R\Pj^2$, both tiled by 12 associahedra $K_4$.
\end{exmp}

\begin{figure}[h]
\includegraphics {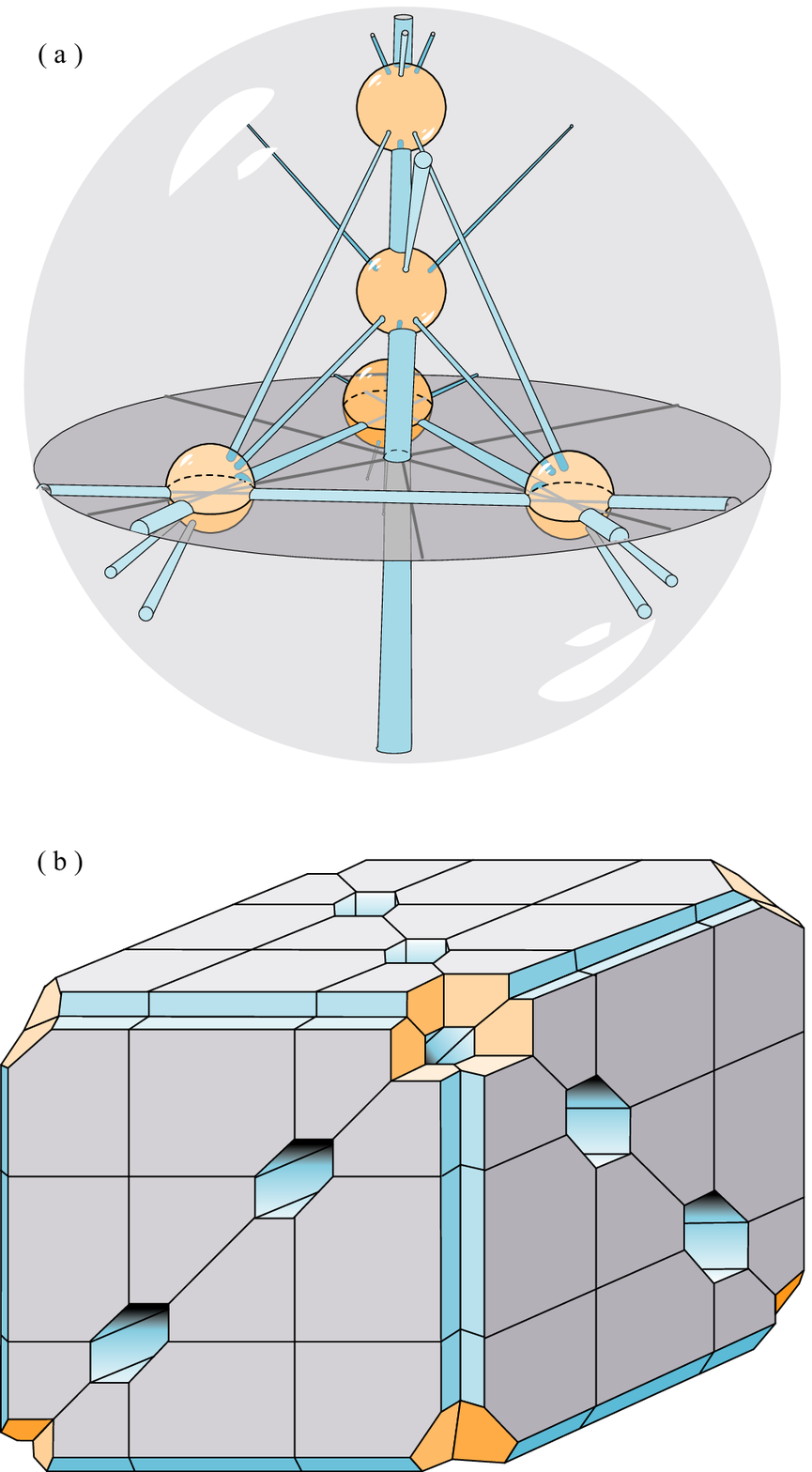}
\caption{Iterated blow-ups of (a) $\R\Pj^3$ to \PVM{3} and (b) $T^3$ to \M{6} are both homeomorphic with a tiling by 60 associahedra.}
\label{pv3m06}
\end{figure}

\begin{exmp}
Figure~\ref{pv3m06}(a) shows $\R\Pj^3$ along with five vertices
(shaded orange) and ten lines (shaded blue) blown-up resulting in \PVM{3}.  All chambers have
been truncated from the simplex to $K_5$. Figure~\ref{pv3m06}(b)
is the blow-up of the $3$-torus into \M{6} along three vertices (orange)
and ten lines (blue). Notice the appearance of the associahedra as in
Figure~\ref{F3-k5}. The resulting manifolds are homeomorphic,
tiled by 60 associahedra. The lower
dimensional moduli spaces \PVM{2} and \M{5} can be seen in the figures
due to a product structure that is inherent in these spaces.
\end{exmp}

\begin{rem}
The iterated blow-up of the minimal building set (that is, the
Fulton-MacPherson compactification) is the key to this
equivalence.  Iterated blow-ups along the {\em maximal} building
set (also known as the {\em polydiagonal} compactification of
Ulyanov), the collection of {\em all} crossings not just the
non-normal ones, yield different manifolds for \PVH{n} and
\oM{n+3}.  For example, the blow-up of \PVH{2} is homeomorphic
to  $\#^8 \, \R \Pj^2$ tiled by 12 hexagons (permutohedra)
whereas \oM{5} is homeomorphic to  $T^2 \, \#^9 \, \R \Pj^2$
tiled by 6 hexagons and 6 octagons.
\end{rem}

%
%
\section*{Conclusion}

Although the motivating ideas of \CM{n} are now classical, the real analog is starting to develop richly.  We have shown \M{n} to be intrinsically related to the braid arrangement, the Coxeter arrangement of type $A_n$.  By looking at other Coxeter groups, an entire array of compactified configuration spaces have recently been studied, generalizing \M{n} from another perspective \cite{small}.  Davis et al.\  \cite[\S5]{djs} have shown these novel moduli spaces to be aspherical, where all the homotopy properties are completely encapsulated in their fundamental groups.   Furthermore, both \M{n} and \PVM{n} have underlying operad structures: The properties of \M{n} are compatible with the operad of planar rooted trees \cite{ksv}, whereas the underlying structure for \PVM{n} is the mosaic operad of hyperbolic polygons \cite{dev}.

This area is highly motivated by other fields, such as string
theory, combinatorics of polytopes, representation theory, and
others.  We think that \M{n} will play a deeper role with future
developments in mathematical physics.  In his {\em Esquisse},
Grothendieck referred to \CM{5} as `un petit joyau'.  By looking
at the real version of these spaces, we see structure determined
by combinatorial tilings, jewels in their own right.

\begin{ack}
We thank  Jim Stasheff for continued encouragement and Mike Carr, Ruth Charney, and Mike Davis for helpful discussions.
\end{ack}

\newpage
%
%

\bibliographystyle{amsplain}

\end{document}